\documentclass[aps,prb,reprint,superscriptaddress]{revtex4-1}

\usepackage{graphicx}
\usepackage{bm}

\begin{document}


\title{Crystallographic features and state stability of the decagonal quasicrystal in the Al-Co-Cu alloy system}


\author{K. Nakayama}
\email[]{k.n@aoni.waseda.jp}
\author{A. Mizutani}
\affiliation{Department of Electronic and Physical Systems, Waseda University, 3-4-1, Okubo, Shinjuku-ku, Tokyo 169-8555, Japan}

\author{Y. Koyama}
\affiliation{Department of Electronic and Physical Systems, Waseda University, 3-4-1, Okubo, Shinjuku-ku, Tokyo 169-8555, Japan}
\affiliation{Kagami Memorial Research Institute for Materials Science and Technology, Waseda University, 2-8-26, Nishiwaseda, Shinjuku-ku, Tokyo 169-0051, Japan}


\date{\today}

\begin{abstract}
In the Al-Co-Cu alloy system, both the decagonal quasicrystal with the space group of $P\overline{10}m2$ and its approximant Al$_{13}$Co$_4$ phase with monoclinic $Cm$ symmetry are present around 20 at.\% Co-10 at.\% Cu. In this study, we examined the crystallographic features of prepared Al-(30-x) at.\% Co-x at.\% Cu samples mainly by transmission electron microscopy in order to make clear the crystallographic relation between the decagonal quasicrystal and the monoclinic Al$_{13}$Co$_4$ structure.  The results revealed a coexistence state consisting of decagonal quasicrystal and approximant Al$_{13}$Co$_4$ regions in Al-20 at.\% Co-10 at.\% Cu alloy samples.  With the help of the coexistence state, the orientation relationship was established between the monoclinic Al$_{13}$Co$_4$ structure and the decagonal quasicrystal.  In the determined relationship, the crystallographic axis in the quasicrystal was found to be parallel to the normal direction of the (010)$_{\rm m}$ plane in the Al$_{13}$Co$_4$ structure, where the subscript m denotes the monoclinic system.  Based on data obtained experimentally, the state stability of the decagonal quasicrystal was also examined in terms of the Hume-Rothery (HR) mechanism on the basis of the nearly-free-electron approximation.  It was found that a model based on the HR mechanism could explain the crystallographic features such as electron diffraction patterns and atomic arrangements found in the decagonal quasicrystal.  In other words, the HR mechanism is most likely appropriate for the stability of the decagonal quasicrystal in the Al-Co-Cu alloy system.

\end{abstract}

\pacs{61.44.Br, 61.66.Dk, 68.37.Lp, 61.05.cp}

\maketitle

\section{Introduction}
Decagonal quasicrystals having one crystallographic axis have been reported to exist as metastable and stable states in Al-based ternary alloy systems such as Al-Co-Cu and Al-Ni-Co systems.\cite{Tsai1989,Tsai1989a,Daulton1992,Grushko1992,Grushko1993c,Grushko1993b,Grushko1993,Grushko1993a,Saitoh1994,Saitoh1996,Hiraga1991,Deloudi2011,Taniguchi2008,Yubuta2014,Tsai1989b,Edagawa1992,Edagawa1994,Ritsch1996,Saitoh1997,Steinhardt1998,Yan1998,Ritsch1998,Abe2000,Abe2000a,Hiraga2001,Deloudi2007,Strutz2009,Strutz2010,Yuhara2011,Hiraga2013}  According to previous studies on quasicrystals in Al-based alloy systems, decagonal quasicrystals can be basically classified into two groups.\cite{Rabson1991,Saito1992}  Decagonal quasicrystals belonging to one group have the space group of centrosymmetric $P10_5/mmc$, and one of the typical examples in this group is the quasicrystal found in Al-20 at.\% Ni-8 at.\% Co.\cite{Ritsch1996,Saitoh1997,Steinhardt1998,Yan1998,Ritsch1998,Abe2000,Abe2000a,Hiraga2001,Deloudi2007,Hiraga2013}  The space group of the other decagonal-quasicrystal group was identified to be $P\overline{10}m2$ with a loss of central symmetry.\cite{Saitoh1994,Saitoh1996,Taniguchi2008,Saito1992}  Of these two groups of decagonal quasicrystals, in this study, we focused on the decagonal-quasicrystal state in the Al-Co-Cu alloy system, which belongs to the latter group with $P\overline{10}m2$ symmetry.  The interesting point to note here is that annealing of the $P\overline{10}m2$ quasicrystal state at higher temperatures leads to a continuous change to the $P10_5/mmc$ state in the Al-Co-Cu alloy system.\cite{Saitoh1996} It is thus likely that there exists a close crystallographic relation between the decagonal quasicrystals with the space groups of $P\overline{10}m2$ and $P10_5/mmc$ in the Al-Co-Cu alloy system. 

It is known that atomic arrangements of decagonal quasicrystals in Al-based alloy systems can be explained in terms of both a two-dimensional quasi-periodic lattice and decagonal atomic-column clusters as a structural basis, just as in the case of usual crystal structures.\cite{Saitoh1996,Hiraga1991,Deloudi2011,Taniguchi2008,Yubuta2014,Saitoh1997,Steinhardt1998,Yan1998,Abe2000,Hiraga2001,Deloudi2007,Strutz2009,Strutz2010,Hiraga2013,Tsuda1993}  Based on previous studies concerning decagonal quasicrystals in Al-Ni-Co and Al-Co-Cu alloys, the structural basis of quasicrystals with the space group of $P\overline{10}m2$ was identified as the fivefold decagonal-column cluster with a size of about 2.0 nm, while a larger column cluster of about 3.2 nm in size was reported to be present in $P10_5/mmc$ quasicrystals.\cite{Hiraga2001}  As for the physical origin of the state stability of decagonal quasicrystals, for instance, the Hume-Rothery (HR) mechanism and the hybridization mechanism have so far been proposed to explain the presence of a pseudogap at the Fermi level.\cite{Tsai2003,Krajci2006,Rogalev2015,Belin-Ferre2004}  Although the HR mechanism is one of the most appropriate candidates, it must be said that the physical origin of the state stability of decagonal quasicrystals in Al-based alloy systems is still an open question. 

It has been reported that there are decagonal quasicrystals with $P\overline{10}m2$ symmetry and its approximant Al$_{13}$Co$_4$ phase with monoclinic symmetry in the Al-Co-Cu ternary alloy system.  Figure~1 shows an Al-rich side of the 1073-K cross section for the Al-Co-Cu phase diagram, which was reported by Grushko.\cite{Grushko1993}
\begin{figure}
\includegraphics[width=8.5cm]{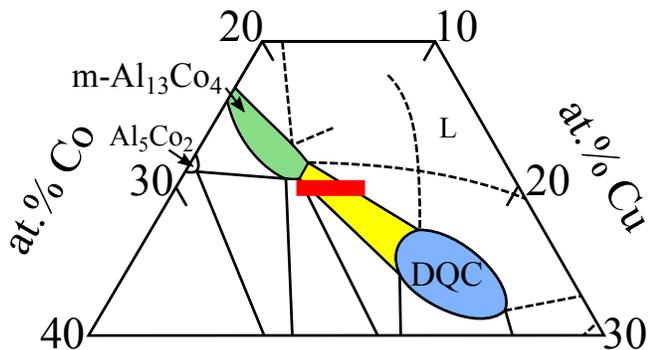}
\caption{(Color online) Al-rich side of the 1073-K cross section in the phase diagram of the Al-Co-Cu alloy system.  The section shown here is a part of the diagram reported by Grushko.\cite{Grushko1993}  In this section, the decagonal-quasicrystal (DQC) and monoclinic Al$_{13}$Co$_4$ (m-Al$_{13}$Co$_4$) phases are present in the blue and green regions, respectively.  There is also a coexistence state consisting of these two phases in the yellow region.  Al-Co-Cu samples with the compositions indicated by the thick red line were used in the experiments to obtain the coexistence state.}
\end{figure}
In the section, the decagonal-quasicrystal state and monoclinic Al$_{13}$Co$_4$ phase are present in the vicinity of the compositions of 17 at.\% Co-17 at.\% Cu and 23 at.\% Co-4 at.\% Cu, respectively.  According to previous studies on intermetallic compounds in the Al-Co-Cu alloy system, the approximant Al$_{13}$Co$_4$ phase has monoclinic $Cm$ symmetry, and the lattice parameters of its unit cell containing 102 atoms were determined to be a = 15.183 \AA, b = 8.122 \AA, c = 12.340 \AA, and β = 107.54$^\circ$.\cite{Hudd1962}  As the decagonal-quasicrystal state and the monoclinic Al$_{13}$Co$_4$ (m-Al$_{13}$Co$_4$) phase are present as two neighboring states in the phase diagram, a coexistence state consisting of these two regions can thus be expected in Al-Co-Cu samples around the composition of 20 at.\% Co-10 at.\% Cu, as indicated by the thick red line in the diagram.  Based on that assumption, the crystallographic features of Al-Co-Cu samples prepared with compositions of Al-(30-x) at.\% Co-x at.\% Cu were examined in this study mainly by transmission electron microscopy in order to make clear the crystallographic relation between atomic arrangements in the decagonal-quasicrystal state and the m-Al$_{13}$Co$_4$ phase.  The determined relationship was used to construct a model based on the HR mechanism.  Concretely, we evaluated the location of the Fermi surface in the decagonal quasicrystal on the basis of the relationship.  The proposed model was then used in an attempt to reproduce the crystallographic features such as electron diffraction patterns and atomic arrangements of the decagonal quasicrystal, which have so far been reported in the Al-Co-Cu alloy system.  In other words, we examined the physical origin of the opening of the pseudogap at the Fermi level in the decagonal quasicrystal in terms of the HR mechanism.

\section{Experimental procedure}
In this study, Al-(30-x) at.\% Co-x at.\% Cu alloy samples with the compositions of 5 $\leq$ x $\leq$ 10 were prepared for use in the experiments.  Ingots of the alloy samples were prepared from Al, Co, and Cu with a purity of 99.99\% by an Ar-arc-melting technique.  To obtain stable states, for instance, alloy ingots were annealed at 1073 K for 24 h for x = 5 and 10 samples, followed by quenching in ice water.  To identify the states appearing in the samples, their x-ray powder diffraction profiles were measured in the angular range of $20^\circ \leq 2\theta \leq 120^\circ$ at room temperature using a RAD-II diffractometer with ${\rm Cu}K\alpha$ radiation.  Both the observation of their microstructures and the determination of the chemical composition in each region were made by means of a JSM-7001F scanning electron microscope equipped with an energy dispersive X-ray spectrometer (EDS).  The crystallographic features of the alloy samples were examined by taking their electron diffraction patterns and corresponding bright- and dark-field images at room temperature using JEM-3010 and 1010 transmission electron microscopes with accelerating voltages of 300 kV and 100 kV, respectively.  Thin specimens for observation by transmission electron microscopy were prepared by using an Ar-ion thinning technique.

\section{Experimental results}
X-ray powder diffraction profiles of the Al-(30-x) at.\% Co-x at.\% Cu samples with 5 $\leq$ x $\leq$ 10 were first measured to identify the states present in them.  It was confirmed that the coexistence state consisting of decagonal-quasicrystal and m-Al$_{13}$Co$_4$ regions was found in Al-20 at.\% Co-10 at.\% Cu alloy samples, while there was basically only the Al$_{13}$Co$_4$ state in Al-25 at.\% Co-5 at.\% Cu ones.  We then focused on the crystallographic features of the former samples in order to identify the crystallographic relation between the decagonal-quasicrystal state and the approximant Al$_{13}$Co$_4$ phase.  An x-ray powder diffraction profile of an Al-20 at.\% Co-10 at.\% Cu sample at room temperature is shown as an example in Fig.~2, together with its backscattered electron image obtained by means of the scanning electron microscope.
\begin{figure}
\includegraphics[width=8.5cm]{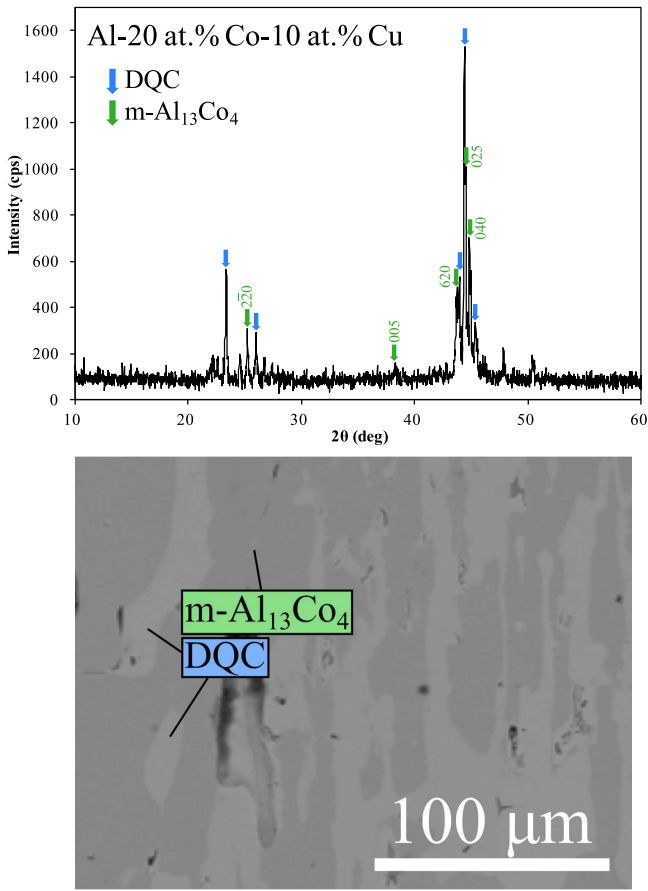}
\caption{(Color online) X-ray powder diffraction profile measured from an Al-20 at.\% Co-10 at.\% Cu sample at room temperature, together with its backscattered electron image obtained by scanning electron microscopy.  The sample was annealed at 1073 K for 24 h as a thermal treatment.  In the profile, stronger reflections are concentrated in the angular range between both $21^\circ \leq 2\theta \leq 27^\circ$ and $43^\circ \leq 2\theta \leq 46^\circ$.  Based on an analysis of these regions, the profile was found to consist of reflections due to the decagonal-quasicrystal state and the m-Al$_{13}$Co$_4$ phase, as indicated by the blue and green arrows.  The image also exhibits two kinds of regions, which give rise to brighter and darker contrasts.  The brighter and darker regions were, respectively, confirmed to be decagonal-quasicrystal and m-Al$_{13}$Co$_4$ regions.}
\end{figure}
In the profile, there are reflections with stronger intensities around $2\theta = 24^\circ$ and $45^\circ$.  Based on a comparison with previously reported profiles of both the states and phases present around the composition of Al-20 at.\% Co-10 at.\% Cu, it was confirmed that the sample consisted of decagonal-quasicrystal and m-Al$_{13}$Co$_4$ regions.  In fact, stronger reflections indicated by the blue and green arrows in the profile were indexed in terms of the decagonal quasicrystal (DQC) and the m-Al$_{13}$Co$_4$ (m-Al$_{13}$Co$_4$) structure, respectively.   In addition, the image exhibits two kinds of regions, which give rise to brighter and darker contrasts.  A distinct feature of the image is that the boundary between these two regions is relatively sharp and smoothly curved.  The EDS analysis also indicated that the brighter- and darker-contrast regions had the compositions of Al$_{68.2}$Co$_{17.9}$Cu$_{13.9}$ and Al$_{72.6}$Co$_{22.3}$Cu$_{5.1}$, respectively.  Based on these experimental data, the states of the brighter and darker regions were identified as the decagonal-quasicrystal state and the m-Al$_{13}$Co$_4$ phase, respectively. 

The coexistence state consisting of decagonal-quasicrystal and approximant Al$_{13}$Co$_4$ regions was present in the samples with the composition of Al-20 at.\% Co-10 at.\% Cu.  These samples were used to examine the crystallographic features of the coexistence state by transmission electron microscopy.  Figure~3 shows a bright-field image taken from one typical area in an Al-20 at.\% Co-10 at.\% Cu sample at room temperature, together with four corresponding electron diffraction patterns.
\begin{figure*}
\includegraphics[width=13.5cm]{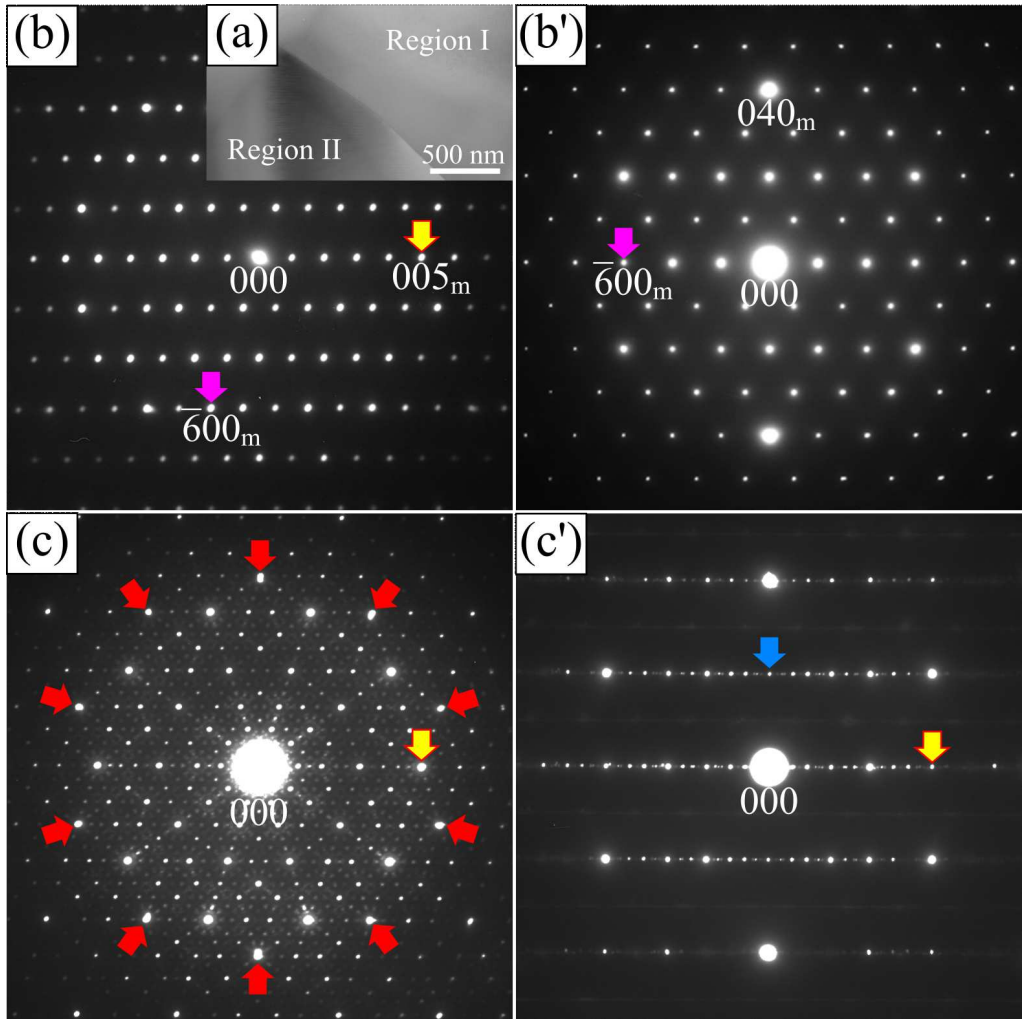}
\caption{(Color online) Bright-field image and four electron diffraction patterns obtained at room temperature, taken from one typical area in an Al-20 at.\% Co-10 at.\% Cu sample annealed at 1073 K for 24 h.  The image exhibits two regions denoted as Regions I and II.   The patterns in (b) and (b') and in (c) and (c') were taken from Regions I and II, respectively.  From an analysis of the patterns in (b) and (b'), Region I was confirmed to be a m-Al$_{13}$Co$_4$ region, and the electron beam incidences are parallel to the normal direction of the (010)$_{\rm m}$ plane for (b) and that of the $(\overline{2}05)_{\rm m}$ plane for (b').  In (b) and (b'), the same $\overline{6}00_{\rm m}$ reflection is also indicated by the purple arrows as an example.  In Region II, ten stronger reflections are arranged with tenfold symmetry in (b), as indicated by the red arrows, and the pattern in (b') exhibits twofold symmetry.  An analysis of experimentally obtained diffraction patterns, including these, indicated that the decagonal-quasicrystal state should appear in Region II.  The location of the reflection indicated by the blue arrow in (c') confirmed that the decagonal-quasicrystal state has a crystallographic periodicity of about 4 nm along the tenfold axis.  Note that the yellow arrows in (c) and (c') indicate the same reflections, which are located at $4 \pi (\sin{\theta / \lambda}) \sim 2.67 {\rm \AA}^{-1}$.}
\end{figure*}
In the image exhibiting a relatively uniform contrast in Fig.~3(a), we can clearly see two regions, which are separated by a curved and sharp boundary.  These two regions are referred to here as Regions I and II.  The notable feature of the image is that a strain contrast is not detected in the vicinity of the boundary.  This implies that Regions I and II should be connected coherently to each other.  To identify the states of Regions I and II, we took electron diffraction patterns of these two regions with various electron beam incidences.  Two diffraction patterns obtained from Region I are shown in Figs.~3(b) and 3(b') as examples, while the patterns in Figs.~3(c) and 3(c') were taken from Region II.  The patterns from Region I exhibit a regular arrangement of reflections, which is an indication that the m-Al$_{13}$Co$_4$ state should be present in this region.  In fact, the reciprocal lattice constructed by using diffraction patterns with various beam incidences was found to be entirely consistent with $Cm$ symmetry for the m-Al$_{13}$Co$_4$ structure as an approximant of the decagonal quasicrystal.  As a result, the electron beam incidences for the patterns in Figs.~3(b) and 3(b') were, respectively, determined to be parallel to the normal directions of the (010)$_{\rm m}$ and ($\overline{2}$05)$_{\rm m}$ planes in the m-Al$_{13}$Co$_4$ structure, where the subscript m denotes the monoclinic system.  In Fig.~3(c) for Region II, on the other hand, ten reflections with stronger intensities are arranged with tenfold symmetry around the origin 000 as the center, as indicated by the red arrows.  Accordingly, the direction of the electron beam incidence for the pattern in Fig.~3(c) is referred to as the tenfold direction, which is parallel to the crystallographic axis.  Based on these features, the decagonal quasicrystal should appear in Region II.  The pattern in Fig.~3(c') can thus be identified as that of the decagonal quasicrystal with the incidence parallel to one of the twofold axes.  The location of the reflection indicated by the blue arrow in Fig.~3(c') implies that the decagonal quasicrystal has a crystallographic periodicity of about 4 nm along the tenfold axis.  These experimental data confirm the actual presence of the coexistence state consisting of decagonal-quasicrystal and m-Al$_{13}$Co$_4$ regions with a coherent boundary in the Al-20 at.\% Co-10 at.\% Cu alloy sample.

In this study, we tried to determine the orientation relationship between the decagonal quasicrystal and the m-Al$_{13}$Co$_4$ structure by using the area shown in Fig.~3(a).  Figure~4 shows an electron diffraction pattern of an area including these two regions, together with its schematic diagram.
\begin{figure}
\includegraphics[width=8.2cm]{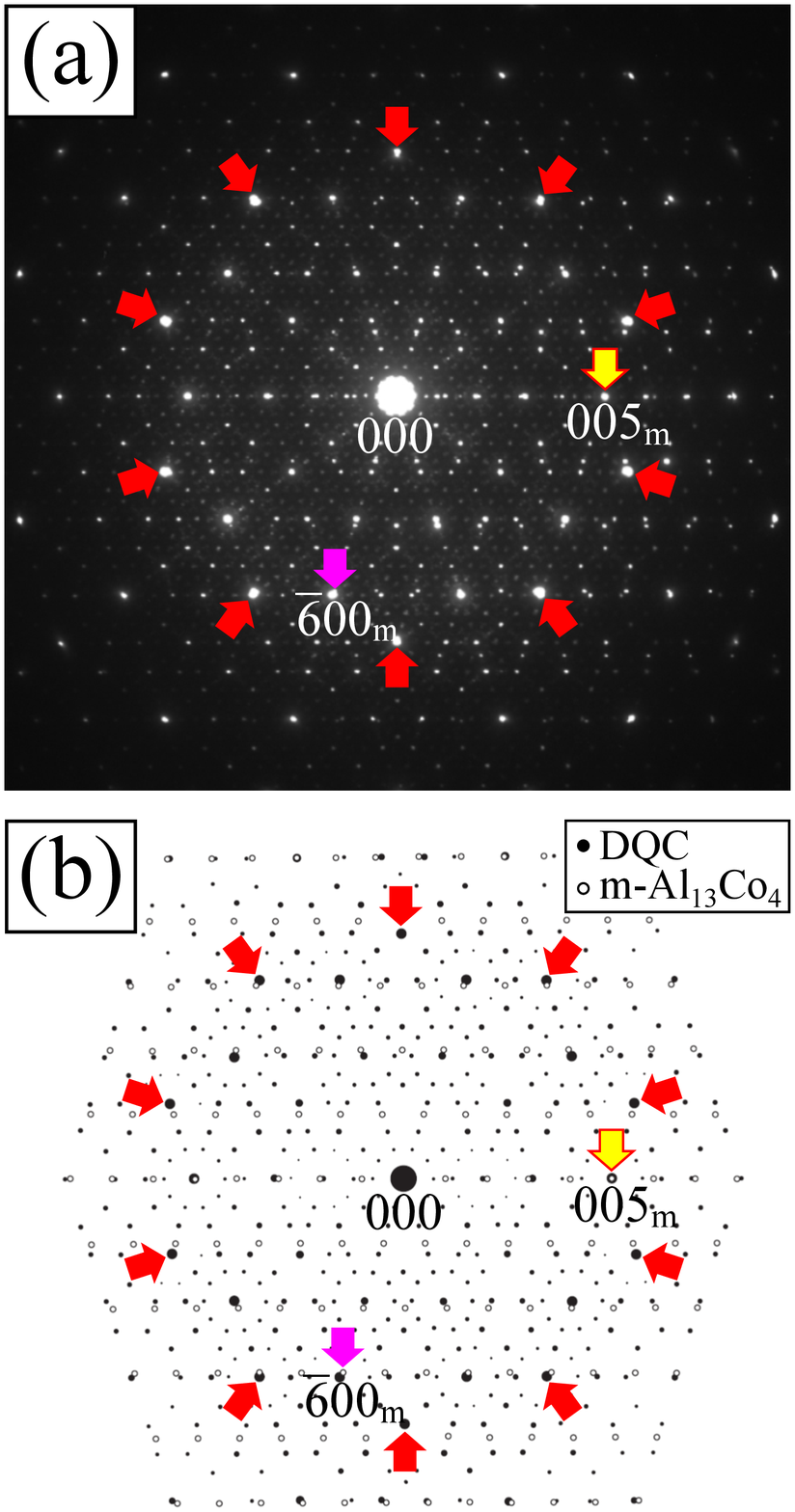}
\caption{(Color online) Electron diffraction pattern of the Al-20 at.\% Co-10 at.\% Cu sample annealed at 1073 K for 24 h, together with its schematic diagram.  The pattern was taken from an area including a decagonal-quasicrystal/m-Al$_{13}$Co$_4$ boundary in the same sample as that used for Fig.~3.  In spite of the involvement of the two regions, the pattern exhibits a relatively simple pattern and is interpreted as being due to a superposition of the pattern of the m-Al$_{13}$Co$_4$ structure with the beam incidence parallel to the normal direction of the (010)$_{\rm m}$ plane and that of the decagonal quasicrystal with the tenfold-axis incidence.  In the diagram, the reflections due to decagonal-quasicrystal and m-Al$_{13}$Co$_4$ regions are indicated by the open and closed circles, respectively.  Note that the red, yellow, and purple arrows in both the pattern and the diagram were already indicated in Fig.~3.  Because the location of the 005$_{\rm m}$ reflection basically coincides with that of one of reflections for the quasicrystal, as indicated by the yellow arrows, the tenfold and twofold axes in the decagonal quasicrystal are parallel to the normal directions of the (010)$_{\rm m}$ and (001)$_{\rm m}$ planes in the m-Al$_{13}$Co$_4$ structure, respectively.}
\end{figure}
In spite of the involvement of the two regions, we can see a relatively simpler pattern in Fig.~4(a).  Its notable feature is that pairs of reflections at the ten positions indicated by the red arrows are likely to be present with pseudo-tenfold symmetry around the origin 000.  We then analyzed the arrangement of reflections in the pattern by comparing them with those in the patterns shown in Fig.~3.  The obtained result is schematically depicted in Fig.~4(b), where the reflections due to the quasicrystal and the m-Al$_{13}$Co$_4$ structure are represented by the open and closed circles, respectively.  Note that the ten red arrows are marked here again, just as in the cases of Figs.~3(c) and 4(a).  The diagram obtained by the analysis revealed that the pattern in Fig.~4(a) could be explained as being due to a superposition of the pattern for the m-Al$_{13}$Co$_4$ structure with the beam incidence parallel to the normal direction of the (010)$_{\rm m}$  plane and that of the decagonal quasicrystal with the incidence parallel to the tenfold axis.  The point to note here is that the location of the 005$_{\rm m}$ reflection for the m-Al$_{13}$Co$_4$ structure is likely to coincide with that of one of reflections for the quasicrystal, as indicated by the yellow arrow.  The value of $4 \pi (\sin{\theta} / \lambda) = 2 \pi / d$ for the 005$_{\rm m}$ reflection was estimated to be about 2.67 \AA$^{-1}$.  Based on these results, we can establish the orientation relationship between the decagonal quasicrystal and the m-Al$_{13}$Co$_4$ structure.  Concretely, the tenfold and twofold axes in the decagonal quasicrystal are, respectively, parallel to the normal directions of the (010)$_{\rm m}$ and (100)$_{\rm m}$ planes in the m-Al$_{13}$Co$_4$ structure. 

\section{Theoretical treatment for the state stability of the decagonal quasicrystal}
The experimental data obtained in this study revealed that there was a clear orientation relationship between the decagonal quasicrystal and the m-Al$_{13}$Co$_4$ structure.  Based on the results obtained experimentally, we propose here a simple model for the state stability of the decagonal quasicrystal; that is, the physical origin of the opening of the pseudogap at the Fermi level in terms of the HR mechanism on the basis of the nearly-free-electron approximation.  In this model, the Fermi sphere with the Fermi radius $k_F$ in reciprocal space was first assumed to be present for both the decagonal quasicrystal and the m-Al$_{13}$Co$_4$ structure, and the radius of the former was derived from that of the latter on the basis of the crystallographic relation between them.  In addition, we assumed that the appearance of a stationary wave with a wave vector of $2\bm{k}_F$ opened the pseudogap at $\bm{k}_F$.  Ten stationary waves in the decagonal quasicrystal were actually adopted to explain the crystallographic features such as its electron diffraction patterns and atomic arrangements.  It should be remarked that the adoption of the ten stationary waves is obviously inconsistent with point-group symmetry, which can be allowed in three-dimensional crystallography.  In other words, it should result in the appearance of atomic bonds, which are shorter than those expected for metallic bonding. 

We start with the estimation of valence-electron concentrations $e/a$ for both the decagonal-quasicrystal state and the m-Al$_{13}$Co$_4$ phase in the Al-20 at.\% Co-10 at.\% Cu samples used in this study.  In our treatment, the values of $e/a$ for Al, Co, and Cu were, respectively, assumed to be +3, -1.66, and +1, which were reported by Raynor.\cite{Raynor1949}  The present EDS analysis also indicated that m-Al$_{13}$Co$_4$ and quasicrystal regions had the chemical compositions of Al$_{68.2}$Co$_{17.9}$Cu$_{13.9}$ and Al$_{72.6}$Co$_{22.3}$Cu$_{5.1}$, respectively.  As a result, the $e/a$ values for the m-Al$_{13}$Co$_4$ phase and the decagonal-quasicrystal state were estimated to be about 1.86 and about 1.89, respectively.  Then, we calculated the Fermi radius of the m-Al$_{13}$Co$_4$ phase by using the simple equation of $k_F^{Al13Co4} = (3 \pi ^2 \rho)^{1/3}$ with $\rho = \frac{n\times(e/a)}{v_c}$, where $n$ is the number of valence electrons involved in a unit cell with a volume $v_c$.  With the help of $n$ = 102, $e/a$ = 1.86, and $v_c$ = 1448 \AA$^3$, we obtained $k_F^{Al13Co4} = 1.57$ \AA$^{-1}$.  As for the Fermi radius of the decagonal quasicrystal, in this treatment, $k_F^{DQC}$ was assumed to be equal to $k_F^{Al13Co4}$ for $e/a$ = 1.89.  We then obtained $k_F^{DQC}=k_F^{Al13Co4} (\frac{189}{186})^{1/3} = 1.58$ \AA$^{-1}$.  It should be noted that this value is quite compatible with that reported recently for the decagonal quasicrystal in the Al-Co-Ni alloy sysytem.\cite{Rogalev2015}

We discuss here the directions of the wave vectors for the ten stationary waves in our model based on the HR mechanism.  Each stationary wave is formed by a superposition of a plane wave with the wave vector $\bm{k}_F$ and its complex-conjugated wave with $-\bm{k}_F$, and the wave number of a stationary wave is twice as large as that of the plane wave.  That is, the stationary wave for the quasicrystal has a wave number of $2k_F$.  As for the direction of the wave vector of each stationary wave, in the case of the decagonal quasicrystal, we need to produce both the two-dimensional quasi-periodicity and the one-dimensional crystallographic periodicity along the tenfold axis at the same time.  In this treatment, we then adopted ten wave vectors with a common crystallographic component, which are schematically depicted in Fig.~5.
\begin{figure}
\includegraphics[width=8.5cm]{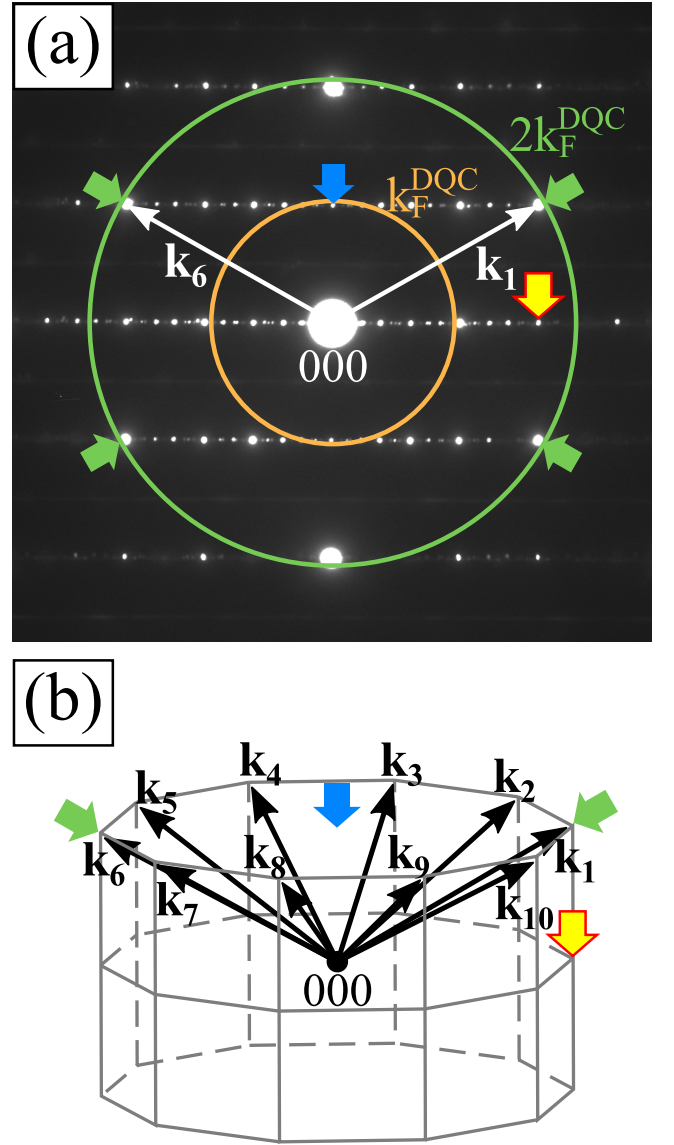}
\caption{(Color online) Electron diffraction pattern of the decagonal quasicrystal with the electron beam incidence parallel to the twofold axis and a three-dimensional diagram of a reciprocal space, indicating the ten wave vectors of the stationary waves with a magnitude of $t\left(2k_F^{DQC}\right)$. The truncation factor $t$ should be associated with the deviation of the Fermi surface from the Fermi sphere, and was estimated to be $t\sim0.976$.  In the (a) pattern, the two vectors denoted as $\bm{k}_1$ and $\bm{k}_6$ are shown, together with the orange and green circles having radii of $k_F^{DQC}$ and $2k_F^{DQC}$. The yellow and blue arrows are shown again, just as in the case of Fig.~3(c').  Among six reflections located near the $2k_F^{DQC}$ circle, the four reflections indicated by the green arrows have quasi-periodic and crystallographic-periodic components.  Because of this, the wave vectors, $\bm{k}_1$ and $\bm{k}_6$, were adopted in our treatment.  As a result, the ten stationary waves adopted here have wave vectors $\bm{k}_j$ with integer $j$ values from 1 to 10, as indicated in the three-dimensional diagram, (b).}
\end{figure}
In Fig.~5(a), for instance, the two wave vectors denoted as $\bm{k}_1$ and $\bm{k}_6$ are shown in the electron diffraction pattern with twofold symmetry, together with two circles with radii of $k_F^{DQC}$ and $2k_F^{DQC}$ in reciprocal space.  Note that the pattern in Fig.~5(a) is identical to that shown in Fig.~3(c').  As seen in the figure, six reflections with stronger intensities are located near the circle for $2k_F^{DQC}= 3.16$ \AA $^{-1}$.  Among these six reflections, the four reflections indicated by the green arrows have wave vectors consisting of quasi-periodic and crystallographic-periodic components.  Taking into account both the presence of these two components and the deviation of their locations from $2k_F^{DQC}$, the $\bm{k}_1$ and $\bm{k}_6$ vectors with $|\bm{k}_1| = |\bm{k}_6| = t\left(2k_F^{DQC}\right)$, not $2k_F^{DQC}$, are adopted in our treatment, where the parameter $t$ is called a truncation factor.\cite{Sato1961}  The factor should be associated with the deviation of the Fermi surface from the Fermi sphere, and was estimated to be $t\sim0.976$ for the present case.  Then, the magnitudes of the quasi-periodic and crystallographic-periodic components of the $\bm{k}_1$ vector are given by $\left[ t\left(2k_F^{DQC}\right) \right] \cos{30^\circ} \sim 2.67$ \AA $^{-1}$ and $\left[ t\left(2k_F^{DQC}\right) \right] \sin{⁡30 ^\circ} \sim 1.54$ \AA $^{-1}$, respectively.  Of these two components, the crystallographic component apparently corresponds to the spacing of $\frac{2\pi}{1.54} \sim 4.08 $ \AA \space for the crystallographic periodicity along the tenfold axis in the decagonal quasicrystal.  To produce tenfold symmetry, we adopted ten wave vectors $\bm{k}_j$ with integer $j$ values from 1 to 10, which are indicated by the thick black arrows in Fig.~5(b).  It should be noted again that these ten wave vectors have a common crystallographic component, which corresponds to a spacing of about 4 \AA \space in real space. 

In our simple treatment, the crystallographic features found in the decagonal quasicrystal are assumed to be the response of a lattice system to the appearance of the above-mentioned stationary waves with the wave vectors $\bm{k}_j$.  To get the stationary wave, we first write an electronic state with a wave vector $\bm{k}_j/2$ as $\phi_j(\bm{r}) = \phi_{0j} \exp{\{i(\bm{k}_j/2) \cdot \bm{r} \}}$ in the plane-wave form with a complex amplitude $\phi_{0j}=|\phi_{0j}|exp(i\theta_j)$.  Note that the wave vector $\bm{k}_j/2$ has the magnitude of $t \left( k_F^{DQC} \right)$.  As was mentioned above, each stationary wave for the S-like state in the nearly-free-electron approximation can be formed by a superposition of the plane wave and its complex-conjugated one.\cite{Ziman1972}  As a result, the stationary wave $\rho_j (\bm{r})$ is expressed by
\begin{equation}
\rho_j (\bm{r}) =  e | \phi_j (\bm{r})+\phi_j^* (\bm{r})|^2 = 4 e |\phi_{0j}|^2 [\cos{\{(\bm{k}_j /2) \cdot \bm{r} + \theta_j \}}]^2,
\end{equation}
where $e$ is the charge of an electron.  Because of
\begin{eqnarray*}
[\cos⁡{\{(\bm{k}_j/2) \cdot \bm{r} + \theta _j\}}]^2 = \{\cos(\bm{k}_j \cdot \bm{r} + 2\theta_j)+1\}/2,
\end{eqnarray*}
the total stationary wave $P(\bm{r})$ becomes
\begin{equation}
P(\bm{r}) = \sum_{j=1}^{10}2e|\phi_{0j}|^2 \{\cos⁡{(\bm{k}_j \cdot \bm{r} + 2 \theta_j)} + 1\}.
\end{equation}
By letting $\psi_j=2\theta_j$, $A_j=2e|\phi_{0j}|^2$, and $B=\sum_{j=1}^{10}A_j$, we can also express $P(\bm{r})$ in the following form, 
\begin{equation}
P(\bm{r}) = \sum_{j=1}^{10}A_j \cos{(\bm{k}_j \cdot \bm{r} + \psi_j)} + B.
\label{Rho}
\end{equation}
As for the response of the lattice system, the screening of the charge induced by the appearance of the total stationary wave should be compensated by the arrangement of atoms (ion cores), which is produced by the lattice waves with the periodicity of $\bm{g}_j=\bm{k}_j$.  Based on this, we can reproduce electron diffraction patterns of the decagonal quasicrystal by using the equation,
\begin{equation}
\bm{G} = \sum_{j=1}^{10} h_j \bm{g}_j,
\label{G}
\end{equation}
with integers $h_j$.  In other words, the ten vectors of $\bm{g}_j$ for $j$ values from 1 to 10 are referred to here as the basic reciprocal lattice vectors for the decagonal quasicrystal.

To check the validity of our theoretical treatment, we first calculated electron diffraction patterns of the decagonal quasicrystal by using Eq.~(\ref{G}).  Three calculated diffraction patterns of the quasicrystal are shown in Fig.~6 as examples.
\begin{figure}
\includegraphics[width=5.8cm]{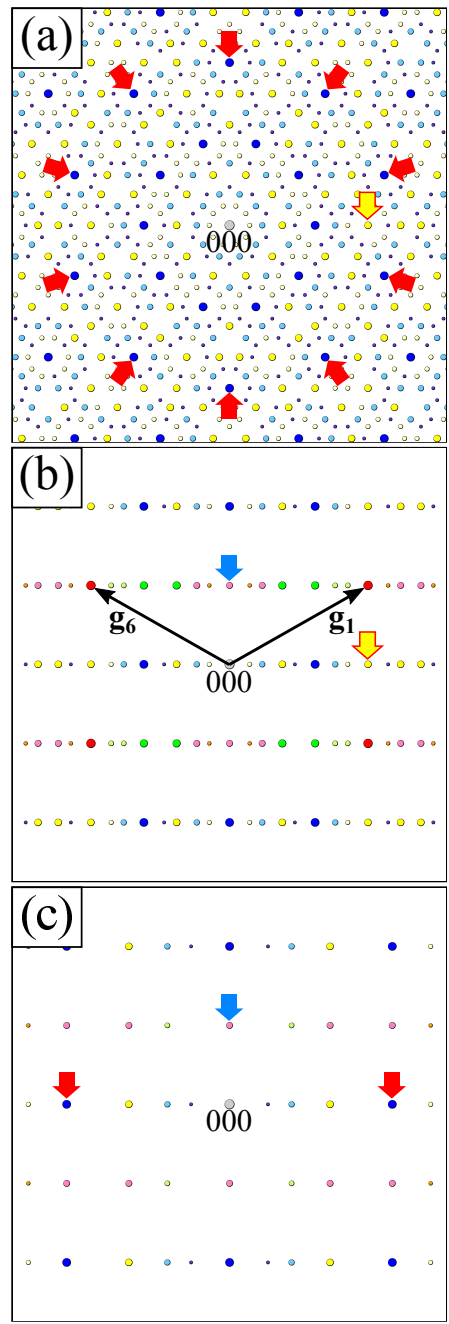}
\caption{(Color online) Schematic diagrams indicating three electron diffraction patterns of the decagonal quasicrystal predicted with our simple theoretical model.  The patterns were actually calculated by using Eq.~(\ref{G}) for integers $h_j$ under the condition of $\sum_{j=1}^{10}|h_j| \leq 10$.  The electron beam incidences for the (a), (b), and (c) patterns are, respectively, parallel to the tenfold axis, one twofold axis, and the other twofold axis.  In the patterns, the reflections with higher $|h_j|$ values are represented by the smaller circles of different colors.  In particular, the first-order reflections for $\bm{k}_j$ are indicated by the red circles.  To easily compare the experimental and calculated patterns, the red, yellow, and blue arrows are also shown in the diagrams.  In spite of the absence of the other reflections with $\sum_{j=1}^{10}|h_j| > 10$, the calculated patterns well reproduce the experimental patterns in Figs.~3(c) and 3(c').}
\end{figure}
Because there are two kinds of twofold axes, the electron beam incidences for the (a), (b), and (c) patterns are, respectively, parallel to the tenfold axis, one twofold axis, and the other twofold axis.  In addition, integers $h_j$ under the condition of $\sum_{j=1}^{10}|h_j| \leq 10$ were used in the calculation, and reflections with higher $|h_j|$ are indicated by the smaller circles of different colors in the patterns.  In spite of the absence of the other reflections with $\sum_{j=1}^{10}|h_j| > 10$, the calculated patterns in Figs.~6(a) and 6(b) are found to well reproduce the experimental patterns shown in Figs.~3(c) and 3(c').  The (c) pattern with the twofold axis is also consistent with that reported previously.\cite{Taniguchi2008}  It is thus suggested that our theoretical treatment should be appropriate for understanding the crystallographic features found in the decagonal quasicrystal.

We recognize that Eq.~(\ref{Rho}) is not a convenient form for the calculation of the charge-density distribution, which is produced by the appearance of the total stationary wave.  In addition, it is hard to understand the correspondence between our treatment and the usual approach in terms of the six-dimensional picture.  To improve these points, we write the total stationary wave in another form, which is derived from Eq.~(\ref{Rho}).  Before showing the final expression, we first introduce both the quantity $\Theta (\bm{r}) = P(\bm{r}) - B$ and the new basic wave vectors $\bm{k}_j$ with $j$ values from 11 to 16, which are shown in Fig.~7.
\begin{figure}
\includegraphics[width=7.5cm]{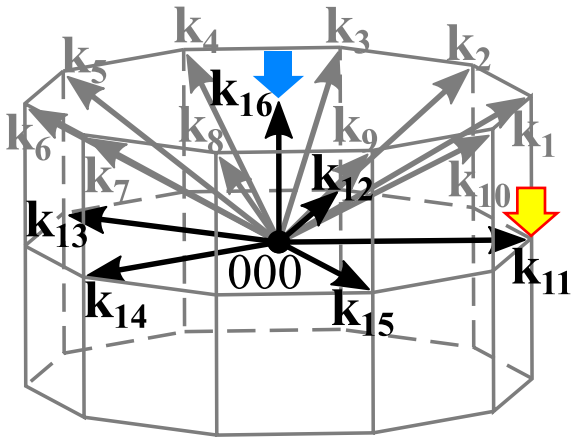}
\caption{(Color online) Schematic diagram indicating another set of wave vectors and also the relation with the wave vectors for the ten stationary waves.  As shown in the diagram, the new set of six wave vectors $\bm{k}_j$ with $j$ values from 11 to 16 was chosen by taking into account the quasi-periodic and crystallographic-periodic components of each wave vector of the stationary wave.}
\end{figure}
The point to note here is that, in the six-dimensional framework using these new vectors, the two-dimensional quasi-periodicity and one-dimensional crystallographic periodicity can formally be treated independently, just as in the case of the usual picture.  Based on the point-group symmetry of the decagonal quasicrystal, the phases $\psi_j$ in Eq.~(\ref{Rho}) are assumed to be $\psi_j=\psi_0 + \Delta$ for $j = 1, 3, 5, 7, 9$ and $\psi_j = \psi_0 - \Delta$ for $j = 2, 4, 6, 8, 10$.  With the help of $A_j = 1$, for simplicity, we obtain the following convenient form:
\begin{equation}
\Theta(\bm{r}) = 2 \left\{ \sum_{j=11}^{15} \cos(\bm{k}_j \cdot \bm{r} + \Delta) \right\} \cos(\bm{k}_{16} \cdot \bm{r} + \psi_0),
\label{Theta}
\end{equation}
for the calculation.  In this study, we actually calculated the spatial distribution of $\Theta(\bm{r})$ as the charge-density distribution for all the stationary waves by using Eq.~(\ref{Theta}), instead of Eq.~(\ref{Rho}).  It should be remarked that Eq.~(\ref{Theta}) contains only two adjustable parameters: the phases of $\Delta$ and $\psi_0$.

As was mentioned earlier, decagonal quasicrystals with $P\overline{10}m2$ and $P10_5/mmc$ symmetries have been reported in the Al-Co-Cu alloy system.\cite{Saitoh1996} These decagonal quasicrystals commonly have a crystallographic periodicity of about 4 \AA \space along the tenfold axis and can be regarded as layered structures consisting of two layers.  In this calculation, the locations of these two layers are specified by $z = 0$ and $z = \frac{1}{2}$ for the assumption of $\psi_0 = 0$ rad.  Figure~8 shows the calculated charge-density distribution $\Theta(\bm{r})$ around the origin indicated by the central black dot for the two cases of $\Delta_1 = 90 \times \frac{\pi}{180}$ rad and $\Delta_2 = 80 \times \frac{\pi}{180}$ rad.
\begin{figure*}
\includegraphics[width=13.5cm]{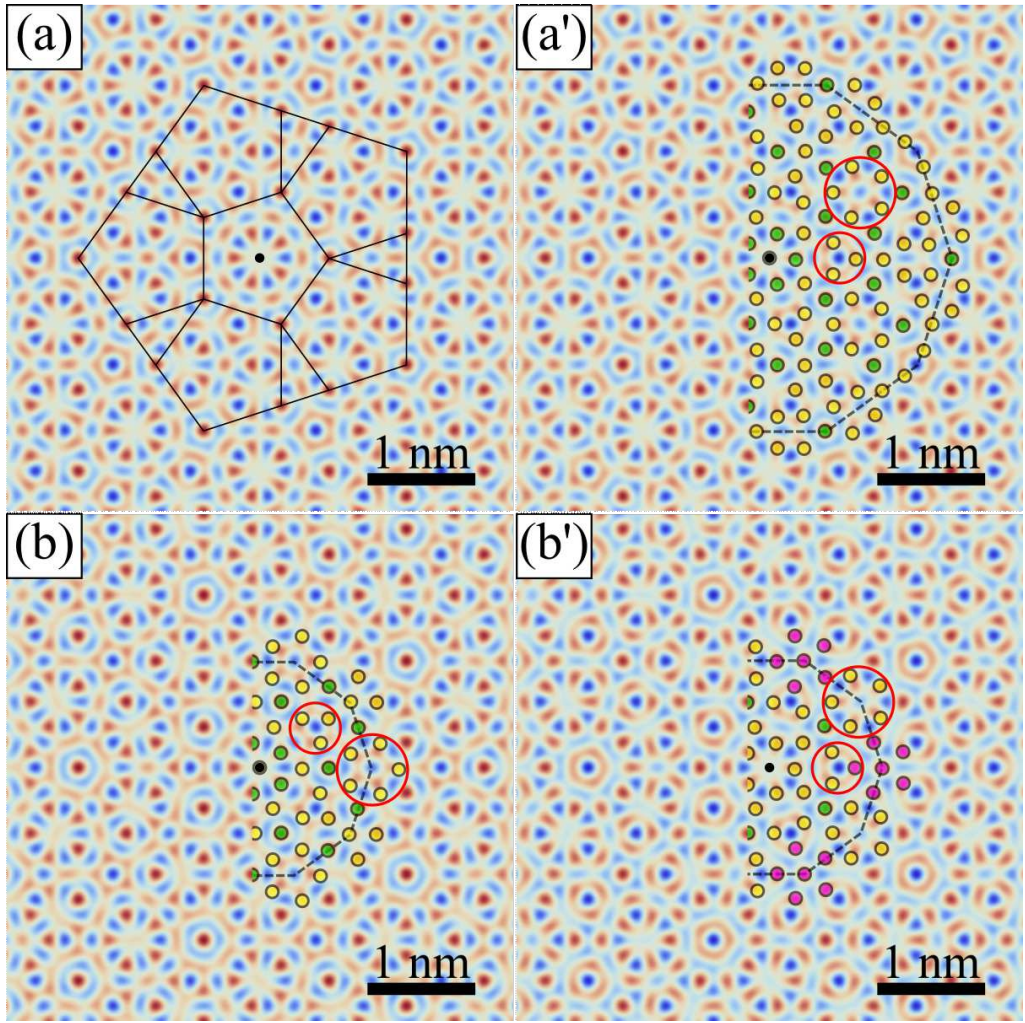}
\caption{(Color online) Charge-density distributions $\Theta(\bm{r})$ at the $z = 0$ and $\frac{1}{2}$ layers in the decagonal quasicrystal, predicted with our model involving two adjustable parameters, $\Delta$ and $\psi_0$.  In addition to the assumption of $\psi_0 = 0$, the calculation was carried out for the two cases of $\Delta_1 = 90 \times \frac{\pi}{180}$ rad for (a) and (a') and $\Delta_2 = 80 \times \frac{\pi}{180}$ rad for (b) and (b').  The calculated distributions around the origins at the $z = 0$ and $\frac{1}{2}$ layers are, respectively, depicted in the diagrams in (a) and (b), and (a') and (b'), where the origin is marked by the central black dot in each diagram.  The magnitude of $\Theta(\bm{r})$ in each diagram decreases in the order of the orange, faint orange, faint blue, and blue colors, and self-similarity can be found in the patterns, as indicated by the black thin lines in (a).  In addition, the atomic positions in the 3.2- and 2.0-nm column clusters, reported by Hiraga and coworkers,\cite{Hiraga2001} are, respectively, depicted in (a'), and (b) and (b'), where transition-metal (TM), Al, and mixed sites are indicated by the green, yellow, and purple circles.  The charge-density distributions predicted with our model are consistent with the reported atomic positions in the column clusters.}
\end{figure*}
The calculated distributions at $z = 0$ and $\frac{1}{2}$ are actually depicted in Figs.~8(a) and 8(a') for $\Delta_1$ and in Figs.~8(b) and 8(b') for $\Delta_2$, respectively.  In the distributions, the magnitude of the $\Theta(\bm{r})$ value decreases in the order of the orange, faint orange, faint blue, and blue colors.  In Figs.~8(a) and 8(a'), first, we can see the characteristic pattern reflecting fivefold symmetry and self-similarity, as indicated by the black lines.  The notable feature of the distributions for $\Delta_1$ is that the pattern at $z = \frac{1}{2}$ can be obtained from that at $z = 0$ by a 36$^\circ$ rotation about the tenfold axis.  This implies that the decagonal quasicrystal with the charge-density distributions at $z = 0$ and $\frac{1}{2}$ should have the space group of $P10_5/mmc$.  As mentioned earlier, the structural basis for the $P10_5/mmc$ quasicrystal in Al-Ni-Co alloys was identified as the 3.2-nm column cluster, which was proposed by Hiraga and coworkers.\cite{Hiraga2001} One-half of the cluster is shown in Fig.~8(a') for $z = \frac{1}{2}$ in order to understand the correspondence between the charge-density distribution and their reported atomic positions in the cluster.  In the schematic diagram of the cluster, the green, yellow, and purple circles represent transition-metal (TM), Al, and mixed sites, respectively.  A comparison indicates that the orange positions with higher densities are occupied by TM atoms, while Al atoms sit on positions with intermediate densities, which are shown by the faint orange color.  The notable feature of the correspondence is that, in the areas surrounded by the red open circles, Al atoms are shifted by a magnitude of about 0.7 \AA \space from the local maximum positions along ridges with respect to the charge density.  From a simple analysis of the distribution, the distance between two neighboring positions with the local maximum density was estimated to be about 2.0 \AA, but the actual distance between Al atoms was determined to be about 2.8 \AA.  It is apparent that the appearance of the shorter distances of about 2.0 \AA \space originates from the adoption of tenfold symmetry in our treatment, which is not allowed in three-dimensional crystallography.  Anyhow, the calculated charge-density distributions are quite compatible with the determined atomic positions in the 3.2-nm column cluster.

In addition to the case of $\Delta_1$, the calculated charge-density distributions $\Theta(\bm{r})$ for the $\Delta_2$ case are shown in Figs.~8(b) and 8(b'), together with schematic diagrams of the reported atomic positions at $z = 0$ and $\frac{1}{2}$ in the 2.0-nm column cluster.  Note that the atomic positions in the 2.0-nm cluster were also reported by Hiraga and coworkers.\cite{Hiraga2001}  It is seen in the figure that the distribution at $z = \frac{1}{2}$ is slightly different from that at $z = 0$.  Concretely, from Eq.~(\ref{Theta}) with $\psi_0 = 0$, the former distribution can be obtained from the latter by a contrast reversal.  This implies that the $10_5$ screw axis is absent in the case of $\Delta_2$ and that the quasicrystal should have $P\overline{10}m2$ symmetry with the 2.0-nm cluster as a structural basis.  As for the correspondence between the distributions and the reported atomic positions, TM and Al atoms sit on the orange and faint orange positions, respectively, just as in the case of $\Delta_1$.  Atomic shifts of Al atoms by a magnitude of about 0.7 \AA \space are also confirmed, as shown in the area surrounded by the red open circles.  Based on these features, the atomic arrangements in the 2.0- and 3.2-nm column clusters are shown to be entirely compatible with the charge-density distributions for $\Theta(\bm{r})$, which were predicted by our simple theoretical model. 

To check the direct correspondence between the charge-density distribution $\Theta(\bm{r})$ and the experimental data, a projection of the distribution along the crystallographic fivefold direction was compared with STEM images reported by Taniguchi and Abe.\cite{Taniguchi2008}  In the comparison, we used the distribution of $|\Theta(\bm{r})|$ as a projected distribution, instead of that of $\Theta(\bm{r})$ itself.  The reason for the adoption of $|\Theta(\bm{r})|$ is that, because the values of $\cos(\bm{k}_{16} \cdot \bm{r})$ in Eq.~(\ref{Theta}) are taken to be 1 for $z = 0$ and -1 for $z = \frac{1}{2}$, the distribution at $z = 0$ has an inverse relation with that at $z = \frac{1}{2}$.  As a result, there is a one-to-one correspondence between the distribution of $|\Theta(\bm{r})|$ and the projected distribution of $\Theta(\bm{r})$.  Figure~9 shows the distribution $|\Theta(\bm{r})|$ actually calculated for the $\Delta_2$ case, together with the contour map derived from the distribution with respect to contrast brightness, which was observed in the STEM image [Fig.~3(b) in Phil. Mag. 88, 1949(2008)].
\begin{figure}
\includegraphics[width=8.5cm]{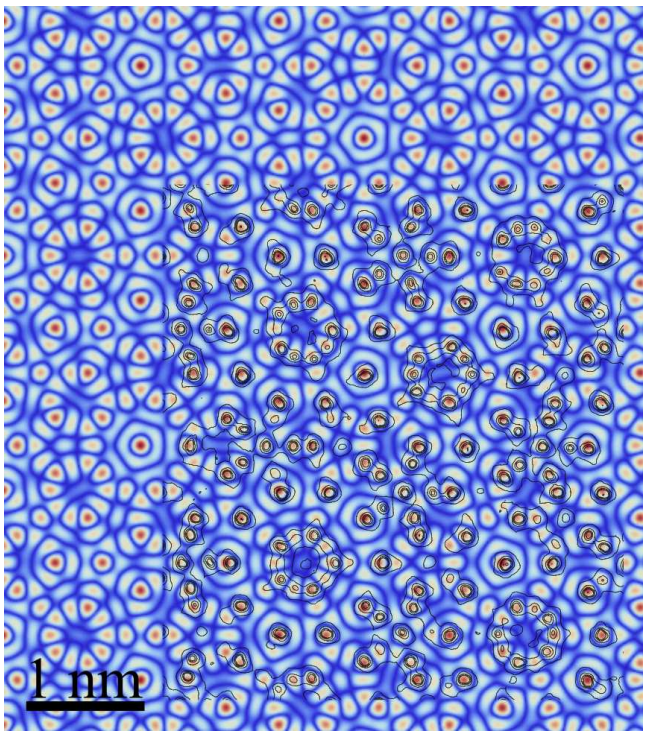}
\caption{(Color online) Schematic diagram indicating the direct correspondence between the predicted charge-density distribution $\Theta(\bm{r})$ and experimental data reported previously.  In the $\Delta_2$ case, the distribution of $|\Theta(\bm{r})|$ was calculated as the projection of the distribution $\Theta(\bm{r})$ along the crystallographic fivefold direction and compared with one of the STEM images reported by Taniguchi and Abe.\cite{Taniguchi2008}  In the distribution, the magnitude of $|\Theta(\bm{r})|$ decreases in the order of the orange, faint orange, faint blue, and dark blue colors.  As for the STEM image, we constructed the contour map from the distribution with respect to contrast brightness in a part of the image [Fig.~3(b) in Phil. Mag. 88, 1949(2008)].  To recover fivefold symmetry in the STEM image, we added the distortion for both dilation of about 4 \% along the horizontal direction and shear of about 0.7$^\circ$. The predicted distribution of $|\Theta(\bm{r})|$ shows excellent agreement with the contour map constructed from the image obtained experimentally.}
\end{figure}
In the calculated distribution, the magnitude of $|\Theta(\bm{r})|$ decreases in the order of the orange, faint orange, faint blue, and dark blue colors.  To recover fivefold symmetry in the STEM image, we also added the distortion imparted to it by both dilation of about 4 \% along the horizontal direction and shear of about 0.7$^\circ$.  The loss of fivefold symmetry in the image is presumably due to experimental difficulties.  As a result, it is seen in the figure that the calculated distribution of $|\Theta(\bm{r})|$ well reproduces the contour map for the STEM image.  In fact, TM and Al atoms were found to occupy the orange and faint orange regions in the calculated distribution, just as in the case of the distributions of $\Theta(\bm{r})$ for $\Delta_1$.  This implies that the spatial distribution of $|\Theta(\bm{r})|$ basically reflects projected atomic positions in the decagonal quasicrystal, which are experimentally obtained by the STEM technique.  This confirms that our theoretical model based on the HR mechanism can well predict the crystllographic features found in decagonal quasicrystals such as their electron diffraction patterns and atomic arrangements.  Therefore, the HR mechanism is probably appropriate for the stability of decagonal quasicrystals with a crystallographic periodicity of about 4 \AA.  This means that the opening of the pseudogap at the Fermi level should originate from the appearance of the stationary waves with $|\bm{k}_j| = t\left(2k_F^{DQC}\right)$.

\section{Discussion}
The present experimental and theoretical results revealed that our simple model based on the HR mechanism can predict the crystallographic features of decagonal quasicrystals with a crystallographic periodicity of about 4 \AA.  In our theoretical model, we assumed the presence of the ten stationary waves with wave vectors $\bm{k_j}$ to explain the opening of the pseudogap at the Fermi level.  The notable feature of our model is the adoption of tenfold symmetry in reciprocal space as a departure from point-group symmetry in three-dimensional crystallography.  As a result, the quasicrystals have to work to avoid a disadvantage factor produced by the adoption of tenfold symmetry.  Here, we will discuss the detailed features of that disadvantage factor on the basis of the calculated distributions $\Theta(\bm{r})$ shown in Fig.~8.

We first pay attention to the distributions of $\Theta(\bm{r})$ at $z = 0$ and $\frac{1}{2}$ shown in the figure.  The correspondence between the distributions and the reported atomic positons in the column clusters indicates that TM atoms occupy the higher-density positions and that Al atoms are shifted by a magnitude of about 0.7 \AA \space from the positons with local charge-density maxima.  The point to note here is that the distance between two neighboring local maxima was estimated to be about 2.0 \AA.  This implies that the adoption of tenfold symmetry results in a shorter distance, which cannot be allowed in real metals and alloys with metallic bonding.  We thus recognize that the appearance of the shorter distance should be regarded as a disadvantage factor produced by the adoption of tenfold symmetry.  The quasicrystals make an effort to avoid the factor through the Al-atomic shifts.  Concretely, the distance between two neighboring Al atoms after the shifts was determined to be about 2.8 Å from the atomic positions in the column clusters, which were reported by Hiraga {\it et al}.\cite{Hiraga2001} The following point should also be noted.  It is known that metallic Al has the face-centered-cubic (fcc) structure with a lattice parameter of 4.0497 Å at 298 K.\cite{Cullity1978}  Accordingly, the shortest distance between two neighboring Al atoms can be estimated to be about 2.8636 Å for metallic bonding.  Based on this, we think that the Al-atomic shifts may be an effort by alloys to avoid the shorter distance as the disadvantage factor produced by the adoption of tenfold symmetry.  In other words, the deviation from the theoretically-predicted atomic positions makes the appearance of the decagonal quasicrystals a reality.  As for the origin of the Al-atomic shifts, it is likely that the usual atomic-size effect is probably the most appropriate candidate, rather than the formation of covalent bonding.

\section{Conclusions}
In this study, we investigated the crystallographic relations between the decagonal quasicrystal and the m-Al$_{13}$Co$_4$ structure in the Al-Co-Cu alloy system mainly by transmission electron microscopy.  Based on experimentally obtained data, we have proposed a simple theoretical model to explain the opening of the pseudogap at the Fermi level in the nearly-free-electron approximation.  In the model, ten stationary waves with the wave-vector magnitude $t\left(2k_F^{DQC}\right)$ were assumed to be present with tenfold symmetry, where the truncation factor $t$ was estimated to be about 0.976.  A comparison with experimental results reported previously showed that the model can predict the crystallographic features of decagonal quasicrystals such as their electron diffraction patterns and atomic arrangements.  On the other hand, the appearance of a shorter distance between two neighboring Al atoms was found to be a disadvantage factor produced by the adoption of tenfold symmetry.  A comparison between the charge-density distribution $\Theta(\bm{r})$ and the reported atomic positions in the column clusters suggested that decagonal quasicrystals tried to avoid the shorter distance through local Al-atomic shifts with a magnitude of about 0.7 \AA.  In other words, the Al-atomic shifts represent an effort by alloys to form decagonal quasicrystals in real space.

\bibliography{refs}

\end{document}